\documentclass[aps,prl,reprint,superscriptaddress,showpacs,amsmath,amssymb,longbibliography,lengthcheck]{revtex4-1}
\usepackage{graphicx}
\usepackage{mathrsfs}
\usepackage{txfonts}
\begin{document}

 \title{Pure Collective Precession Motion of High-Spin Torus Isomer} 

\author{T. Ichikawa}%
\affiliation{Yukawa Institute for Theoretical Physics, Kyoto University,
Kyoto 606-8502, Japan}
\author{K. Matsuyanagi}
\affiliation{Yukawa Institute for Theoretical Physics, Kyoto University,
Kyoto 606-8502, Japan}
\affiliation{RIKEN Nishina Center, Wako 351-0198, Japan}
\author{J. A. Maruhn}
\affiliation{Institut fuer Theoretische Physik, 
Universitaet Frankfurt, D-60438 Frankfurt, Germany}
\author{N. Itagaki}
\affiliation{Yukawa Institute for Theoretical Physics, Kyoto University,
Kyoto 606-8502, Japan}
\date{\today}

\begin{abstract}
 We investigate the precession motion of the exotic torus configuration
 in high-spin excited states of $^{40}$Ca. For this aim, we use the
 three-dimensional time-dependent Hartree-Fock (TDHF) method.  Although
 the high-spin torus isomer is a unique quantum object characterized by
 the alignment of angular momenta of independent single-particle
 motions, we find that the obtained moment of inertia for rotations
 about an axis perpendicular to the symmetry axis is close to the
 rigid-body value. We also analyze the microscopic structure of the
 precession motion using the random-phase approximation (RPA) method for
 high-spin states. In the RPA calculation, the precession motion of the
 torus isomer is generated by coherent superposition of many
 one-particle-one-hole excitations across the sloping Fermi surface that
 strongly violates the time-reversal symmetry. By comparing results of
 the TDHF and the RPA calculations, we find that the precession motion
 obtained by the TDHF calculation is a pure collective motion well
 decoupled from other collective modes.
 \end{abstract}

\pacs{21.60.Jz, 21.60.Ev, 27.40.+z}
\keywords{}

\maketitle

Nuclear rotation is a collective motion that restores the symmetry
spontaneously broken in the self-consistent mean field.  When deformed
nuclei break the spherical symmetry but preserve the axial symmetry, the
rotation about the symmetry ($z$) axis is quantum mechanically
forbidden.  For instance, in high-spin oblate isomer states, the total
angular momentum about the symmetry axis is constructed not by the
collective rotation but the alignments of the angular momenta of
individual nucleons \cite{BM77,Afan99}.  However, even such a state can
rotate about an axis perpendicular to the symmetry axis, because the
density distribution breaks the rotational symmetry about that axis.
Below we call this ($x$ or $y$) axis a perpendicular axis.  This
rotational degree of freedom causes the {\it precession} motion of the
system as a whole \cite{BM81}.

In our previous paper \cite{ich12}, we showed the existence of a stable
torus configuration in high-spin excited states of $^{40}$Ca, whose $z$
component of the total angular momentum, $K$, is $K=60$ $\hbar$.  This
large angular momentum is generated by alignment of single-particle
angular momenta of totally twelve nucleons; the $z$ components of the
orbital angular momenta, $\Lambda=+4$, $+5$, and $+6$ $\hbar$ for
spin-up or -down neutrons and protons, are summed up to $K=60$ $\hbar$.
Thus, this torus isomer has a significant amount of circulating current.
A question then arises how such a ``femto-scale magnet'' rotates
collectively to restore the broken symmetry about a perpendicular axis.

A key physical quantity to understand fundamental properties of nuclear
rotation is the moment of inertia about a perpendicular axis.  It is
theoretically known that an independent-particle configuration in a
deformed harmonic-oscillator potential rotates with rigid moment of
inertia, provided that the self-consistency between the potential and
the density distribution is fulfilled \cite{BMbook}.  In reality,
however, measured moments of inertia for deformed nuclei largely deviate
from the rigid-body values.  For instance, measured moments of inertia
for precession motions of high-$K$ prolate isomers are significantly
smaller than rigid-body values \cite{del04, shi05}.  This reduction has
been seen at high spin where pairing correlations are negligible and is
attributed to shell effects \cite{del04}.  For high-$K$ oblate isomers,
precession modes have not yet been observed. A possible reason is that
their moments of inertia are much reduced from rigid-body values due to
oblate shell structure at small deformation \cite{and81}.  Then
excitation energies of precession motion become higher.  This would be a
 reason why the search for precession modes of high-$K$ oblate
isomer is difficult and remains as an experimental challenge.  The
high-$K$ torus isomer can be regarded as an extreme limit of the
high-$K$ oblate isomer.  Therefore, dynamical properties of the high-$K$
torus isomer revealed in its moment of inertia about a perpendicular
axis will provide a fresh insight into dynamical properties of high-$K$
oblate isomers as well.

In this Letter, we present a periodic numerical solution of the
precession motion of the high-$K$ torus isomer in $^{40}$Ca described by
the three-dimensional time-dependent Hartree-Fock (TDHF) equation.  We
trigger the precession motion by applying a certain amount of angular
momentum in the direction of a perpendicular axis.  We estimate the
moment of inertia characterizing such an exotic mode of nuclear
collective rotation.  We find that the obtained moment of inertia is
close to the rigid-body value.  This result is surprising, because the
high-$K$ torus isomer is created by aligned angular momenta of
independent particle motion and possesses strong time-odd components in
the self-consistent mean field.  We shall also analyze the microscopic
structure of the precession motion using the random-phase approximation
(RPA) method and compare with the result of the TDHF calculation.

Since the TDHF method describes time-evolution of a wave packet,
quantization is necessary to obtain quantum spectra.  If we succeed in
obtaining periodic numerical solutions in real-time evolution of the TDHF
mean field, then we can adopt the semi-classical quantization procedure.
It is, however, very difficult to obtain the periodic solutions, because
nonlinear effects tend to destroy the periodic motion and lead to
chaotic motion.  Thus, up to now, periodic solutions have been found
only for a few relatively simple cases such as large-amplitude monopole
vibrations in $^{4}$He and $^{16}$O \cite{wu99}.  Periodic solutions for
rotational modes have not yet been reported.  In this Letter, we will
also show that the precession frequency obtained in the TDHF calculation
agrees with that of the RPA method in good approximation.  That is, we
have, for the first time, succeeded in obtaining the quantum excitation
energy of the precession motion by the numerical application of the TDHF
method.

\begin{figure}[t]
\includegraphics[keepaspectratio,width=4.5cm]{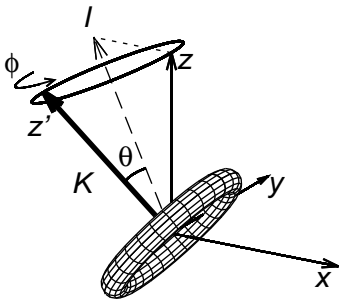}
\caption{Schematic
picture of the precession motion of a high-$K$ torus isomer.  The bold
solid arrow denotes the symmetry axis of the density distribution.  The
dashed arrow denotes the precession axis. The symbols $\theta$ and
$\phi$ denote the tilting and the rotation angles, respectively.}
\end{figure}

\begin{figure*}[htbp]
\includegraphics[keepaspectratio,width=\linewidth]{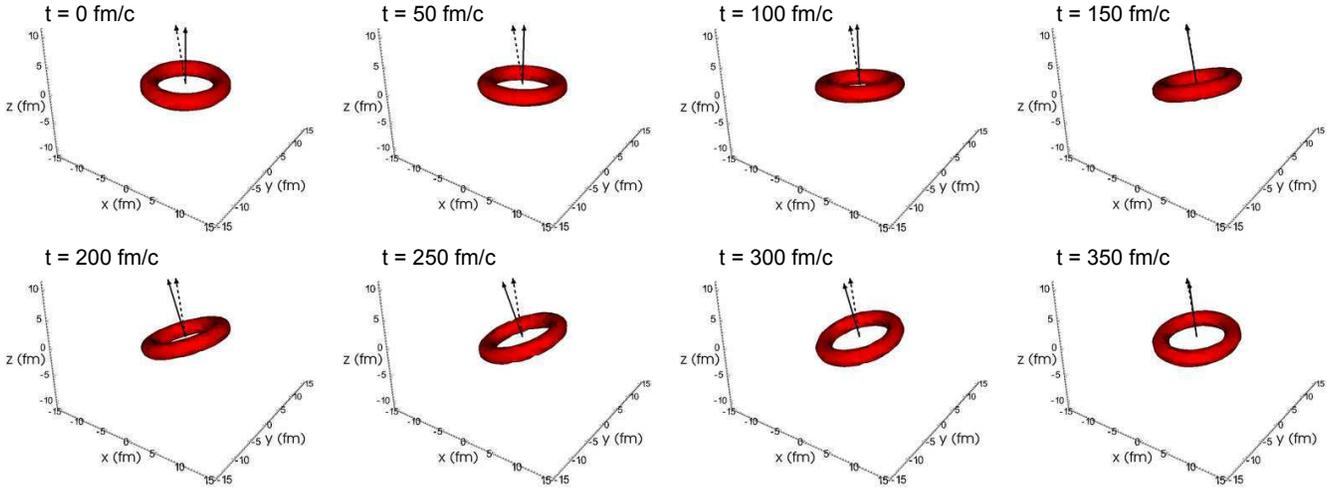}
 \caption{(Color
online) Snapshots of the time evolution of the density distribution of
the high-$K$ torus isomer in $^{40}$Ca obtained by the TDHF
calculations. The (red) surface indicate that the density is half of the
maximum value there. The time step of each snapshot is 50 fm/$c$.
 The solid and the dotted lines denote the symmetry and the precession axes,
 respectively.}
\end{figure*}

Figure 1 shows a schematic picture for the precession motion of a
high-$K$ torus isomer.  In the figure, the bold solid arrow denotes the
symmetry ($z'$) axis of the density distribution in the body-fixed
frame.  At $t=0$, this axis is identical to the $z$ axis in the
laboratory frame and the torus isomer has the angular momentum, $K$,
along this axis.  When we give an angular momentum to the (negative)
direction of the $x$ axis (the dotted line) at $t=0$, the total angular
momentum changes to $\vec I$. Then, the precession motion starts.  The
symmetry ($z'$) axis rotates about a fixed axis (the dashed arrow) that
coincides with the direction of the total angular momentum $\vec I$.  We
call this axis `the precession axis'.  In the precession motion, the
value $K$ is conserved. The tilting angle, $\theta$, is defined as the angle
between the symmetry ($z'$) axis and the precession axis (the direction
of the total angular momentum).  The symbol $\phi$ denotes the rotation
angle of the $z'$ axis rotating about the precession axis.  The moment
of inertia for the rotation about a perpendicular axis of the torus
configuration, $\mathscr{T}_\perp$, is then given by
$\mathscr{T}_\perp=I/\omega_{\rm{prec}}$, where $\omega_{\rm{prec}}$
denotes the rotational frequency of the precession motion.  The
first excited state of the precession motion is the state with $I=K+1$.
Since the torus isomer of $^{40}$Ca has $K=60$ $\hbar$, we calculate the
precession motion with $I=61$ $\hbar$.

To calculate the precession motion by means of the TDHF method, we use
the code {\tt Sky3D} \cite{sky3d}. We calculate the initial state of the torus
configuration for $^{40}$Ca with the $z$ component of the total angular
momentum $K=60$ $\hbar$ by the cranked HF method using this code. The
details are given in Ref.~\cite{ich12}.
In the calculations, the single-particle wave functions are described on
a Cartesian grid with a grid spacing of 1.0 fm. We take
$32\times32\times24$ grid points for the x, y, and z directions,
respectively.  In all the calculations, we use the SLy6 interaction
\cite{Cha97a}.  As shown in Ref.~\cite{ich12}, the interaction
dependence is negligible, because the spin-orbit force effects are weak in the
torus configuration.  The obtained density distribution on the plane at
$z=0$ is well fitted by $\rho(r)=\rho_0e^{-(r-R_0)^2/\sigma^2}$, where
$\rho_0=0.12$ fm$^{-3}$, $R_0=6.06$ fm, and $\sigma=1.64$ fm.  The
rigid-body moment of inertia calculated using the obtained density
distribution is $\mathscr{T}_\perp^{\rm rid}=$ 21.1 $\hbar^2$/MeV.

To excite the precession motion, we provide an impulsive force at $t=0$
fm/$c$ by the external potential given by $V_{\rm ext}(r,\varphi,z)=V_0
z \cos \varphi \exp[-(r-R_0)^2/\sigma^2]$. The parameter $V_0$ is
chosen such that the total angular momentum becomes $I=61$ $\hbar$, that
is, the $x$ component of the total angular momentum, $I_x$, is
$I_x=-11.0$ $\hbar$ at $t=0$ [$I=\sqrt{60^2+(-11)^2}$ $\hbar$ =61
$\hbar$].  Here, we use $V_0=0.12757$ MeV. We determine the time
evolution of the density distribution by solving the TDHF equation of
motion, $i\hbar\dot{\rho}=[h,\rho]$, where $h$ is the single-particle
Hamiltonian and $\rho$ is the one-body density matrix.
To solve the TDHF equation, we take Taylor expansion to the
time-development operator up to the 12 order in the code.
The time step of the
TDHF calculations is 0.2 fm/$c$.  We calculate the time-evolution until
3000 fm/$c$. Thus, we obtain about 7.5 periods of the precession motion.

Figure 2 shows snapshots of the time-evolution of the density
distribution obtained by the TDHF calculations.  In the figure, we plot
the surface at the half of the maximum value of the density
distribution. The time step of each snapshot is 50 fm/c.  We can clearly
see in this figure about one period of the precession motion of the
high-$K$ torus isomer of $^{40}$Ca.

\begin{figure}[htbp]
\includegraphics[keepaspectratio,width=\linewidth]{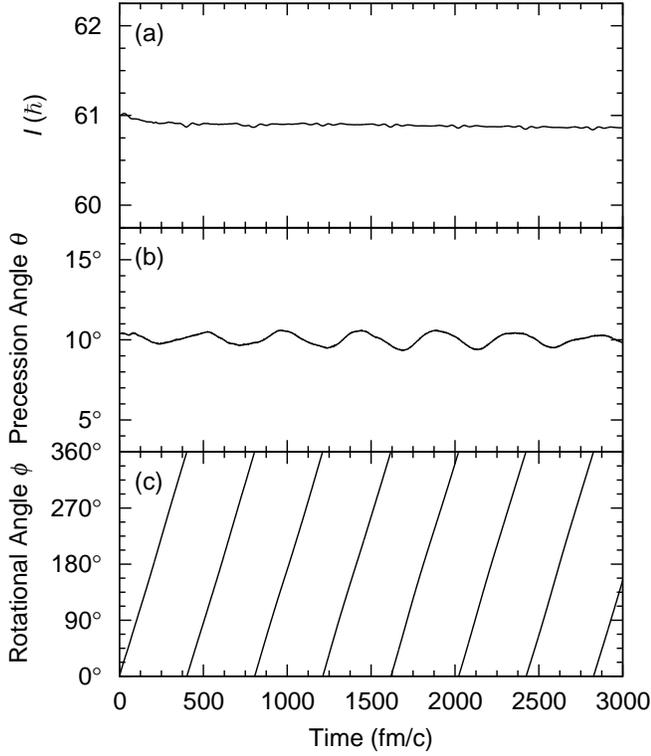}\\%
\caption{Time-evolution of (a) total angular momentum $I$, (b) the
 tilting angle $\theta$, and (c) the rotational angle $\phi$ of the
 precession motion for the high-$K$ torus isomer in $^{40}$Ca.}
\end{figure}

Figure 3 shows the time-evolution of (a) the total angular momentum $I$, (b)
the tilting angle $\theta$, and (c) the rotational angle $\phi$.  In
Fig.~3 (a), we see that the value of the total angular momentum
converges to about 61 $\hbar$, indicating that the TDHF calculations
work well for long duration.  The total energy also converges well.  By
the impulsive force, $-\partial V_{\rm ext}/\partial z$, exerted at
$t=0$, not only the precession motion but also other collective motions
such as the $\gamma$ vibrations might be excited.  However, the tilting
angle fluctuates only slightly between 10$^\circ$ and 11$^\circ$ [see
Fig.~3 (b)] indicating that the coupling effects between the precession
motion and other collective modes are rather weak.  In Fig.~3 (c), we
see that the rotational angle linearly increases in each period. The
obtained periods are 401.4, 403.5, 404.6, 405.4, 403.5, 400.9, and
401.5 fm/$c$.  The fluctuations of the period indicate the extent of the
effects due to the couplings with other collective modes and/or
precision of numerical calculation.  They are much smaller (less than 1
$\%$) than the time scale of the precession motion. The average period,
$T_{\rm prec}$, is 403.0 fm/$c$.  Thus, the average frequency is
$\omega_{\rm prec}=2\pi/T_{\rm prec}=3.08$ MeV$/\hbar$.  We can identify
$\hbar \omega_{\rm prec}$ with the $\Delta I=1$ excitation energy of the
precession mode of the high-$K$ torus isomer.  We shall later confirm
this interpretation in connection with the RPA treatment of this mode,
The moment of inertia obtained in this way is $\mathscr{T}_\perp^{\rm
TDHF}=I/\omega_{\rm prec}=19.8$ $\hbar^2/$MeV, which is very close to
the rigid-body value $\mathscr{T}_\perp^{\rm rid}=21.1$ $\hbar^2/$MeV.
The high-$K$ torus isomer is a unique quantum object characterized by
the alignment of angular momenta of independent single-particle motions.
The alignment causes a significant amount of circulating current and, as a
consequence, the self-consistent mean field strongly violates the
time-reversal symmetry.  Although these features are essentially
different from the classical rigid body, we find that the high-$K$ torus
isomer performs a collective rotation about a perpendicular axis with
the moment of inertia close to the rigid-body value.

To obtain a deeper understanding of the microscopic particle-hole
structure generating the collective precession motion, we have performed
an RPA calculation with the radial displaced harmonic-oscillator (RDHO)
model \cite{Wong73}.  We also confirm the validity of the relation
$\mathscr{T}_\perp^{\rm TDHF}=I/\omega_{\rm prec}$ used to extract the
moment of inertia from the real-time TDHF evolution.  The RDHO model
represents the major features of the torus isomer and works well,
because effects of the spin-orbit force are negligible in the torus
configuration.  We can also avoid the complications for the treatment of
unbound single-particle states by using this model.  In the RDHO model
for the torus configuration, the single-particle potential $V$ is given,
in the cylindrical coordinates, by
$V(r,z)=\frac{1}{2}m\omega_0^2(r-R_0)^2+\frac{1}{2}m\omega_0^2 z^2$,
where $m$ and $R_0$ denote the nucleon mass and the torus radius,
respectively.  The Coulomb potential is ignored for simplicity.  The
harmonic-oscillator frequency $\omega_0$ is determined such that the
density distribution calculated with this model agrees, in good
approximation, with that of the torus isomer obtained by the cranked HF
calculation.  The rigid-body moment of inertia calculated with the RDHO
density distribution is $\mathscr{T}_\perp^{\rm RDHO}=$ 21.1
$\hbar/$MeV, which agrees with the rigid-body value,
$\mathscr{T}_\perp^{\rm rid}$, for the torus isomer.

We can easily determine the frequency of the precession motion by
solving the RPA dispersion equation \cite{shi05},
$\mathscr{T}_\perp^{\rm RPA}(\omega)=K/\omega$, where the moment of
inertia, $\mathscr{T}_\perp^{\rm RPA}$, is a function of $\omega$
defined by
\begin{equation}
\mathscr{T}_\perp^{\rm RPA}(\omega)
=\frac{\hbar^2}{2}\sum_{ph}\left\{\frac{|J_{+}^{ph}|^2}{(\epsilon_{ph}-\hbar\omega)}+
\frac{|J_{-}^{ph}|^2}{(\epsilon_{ph}+\hbar\omega)}\right\}.   
\end{equation}
Here the sum is taken over the one-particle (1p)-one-hole (1h)
excitations across the sloping Fermi surface (see Fig.~3 in
Ref.~\cite{ich12}) and $\epsilon_{ph}$ denote their excitation energies.
The quantities $J_{\pm}^{ph}$ represent the matrix elements of the
angular momentum raising and lowering operators,
$J_{\pm}^{ph}=\left<ph\right|\hat{J_\pm}\left|0\right>$, between the
torus configuration $\left|0\right>$ and the 1p-1h excited states
$\left|ph\right>$.  This RPA dispersion equation is valid for
velocity-independent residual interactions, and it yields the classical
relation, $\mathscr{T}_\perp^{\rm rid}=I/\omega$, except that anharmonic
effects higher order in $1/K$ are ignored and, accordingly, $I$ is
approximated by $K$.  By solving the RPA dispersion equation, we can
simultaneously determine the frequency, $\omega$, and the moment of
inertia, $\mathscr{T}_\perp$, of the precession motion.  The lowest
eigen-frequency that satisfies the above equation is just the RPA
precession frequency of interest.  We denote this solution $\omega_{\rm
RPA}$. The precession motion is generated by coherent
superposition of many 1p-1h excitations across the sloping Fermi
surface. The value of $\mathscr{T}_\perp^{\rm RPA}$ at $\omega_{\rm RPA}$
is the RPA moment of inertia for the precession motion.  In the limit
$\omega_{\rm RPA}=0$, $\mathscr{T}_\perp^{\rm RPA}$ reduces to the
adiabatic cranking formula, $\mathscr{T}_\perp^{\rm crank}$.  Using the
single-particle wave functions obtained by the RDHO model, we determine
$\omega_{\rm RPA}$ and $\mathscr{T}_\perp^{\rm RPA}$.  In the
calculations, we take all 1p-1h excitations whose energies are below
$\epsilon_{ph}\le30$ MeV.  We obtain $\omega_{\rm RPA}=3.02$ MeV/$\hbar$
and $\mathscr{T}_\perp^{\rm RPA}=19.6$ $\hbar^2$/MeV.  This value of the
RPA moment of inertia is different from the adiabatic cranking value
$\mathscr{T}_\perp^{\rm crank}=20.0$ $\hbar^2$/MeV only slightly,
indicating that the effect of the finite frequency ($\omega_{\rm RPA}
\ne 0$) is rather small for the precession motion under consideration.

The RPA frequency $\omega_{\rm RPA}$ and the moment of inertia
$\mathscr{T}_\perp^{\rm RPA}$ agree with the TDHF results for
$\omega_{prec}$ and $\mathscr{T}_\perp^{\rm TDHF}$ in very good
approximation.  If $K$ is replaced with $I=K+1$ in the RPA dispersion
equation, the agreement becomes even better ($\omega_{\rm RPA}=3.07$
MeV/$\hbar$).  This almost perfect agreement clearly indicates that the
periodic numerical solution obtained in the real-time TDHF evolution
describes the collective rotation well decoupled from other collective
modes.   The agreement between the TDHF and RPA results furthermore
suggests that the net effect of the velocity-dependent interactions such
as the spin-orbit interaction is small, despite the presence of
a significant amount of circulating current which strongly violates the
time-reversal symmetry in the self-consistent mean field.  As we have
seen in \cite{ich12}, the effect of the spin-orbit potential almost
cancel out between the inside and outside of the torus radius $R_0$.
This suggests that the velocity-dependent interaction effects are almost
canceled out for the precession motion under consideration.  The results
of the TDHF and the RPA calculations thus suggest that basic physical
conditions for the occurrence of the rigid precession motion are 1) the
independent-particle configuration is pure and stable, 2) the symmetry
breaking about a perpendicular axis is sufficiently strong, and 3) the
net effect of the velocity-dependent interactions is small.
  
In summary, we have obtained a periodic numerical solution in the TDHF
time-evolution, that describes the precession motion of the high-$K$
torus isomer with $K=60$ $\hbar$ in $^{40}$Ca. Although the high-$K$
torus isomer is a unique quantum object characterized by the alignment
of angular momenta of independent single-particle motions, we find that
the torus isomer performs a collective rotation about an axis
perpendicular to the symmetry axis with the moment of inertia close to
the classical rigid-body value.  We have also performed the RPA
calculation for the precession motion with the RDHO model.  By comparing
the results of the TDHF and the RPA calculations, we have confirmed that
the periodic TDHF solution corresponds to the precession mode generated
by coherent superposition of many 1p-1h excitations across the sloping
Fermi surface.  This exotic mode of rotation at high spin is ideally
decoupled from other collective modes.

\begin{acknowledgments}
 A part of this research has been funded by MEXT HPCI STRATEGIC PROGRAM.
  This work was undertaken as part by the Yukawa International Project
  for Quark-Hadron Sciences (YIPQS).  J.A. M. was 
  supported by BMBF under contract number 06FY9086 and 05P12RFFTG,
  respectively.
 \end{acknowledgments}


\begin{thebibliography}{50}
 \bibitem{BM77}
	 A. Bohr, Proc. Int. Physics `Enrico Fermi', Cours LXIX,
	 ed.  A. Bohr and R. A. Broglia (Amsterdam: North-Holland, 1977), p3.
 \bibitem{Afan99}
	 A.V. Afanasjev {\it et al.}, 
	 Phys. Rep. {\bf 322}, 1 (1999).
 \bibitem{BM81}
	 A. Bohr and B.R. Mottelson, Nucl. Phys. {\bf A354}, 303c
	 (1981).
 \bibitem{ich12}
	 T. Ichikawa {\it et al.}, Phys. Rev. Lett. {\bf 109}, 232503 (2012).
 \bibitem{BMbook}
	 A. Bohr and B.R. Mottelson, ``Nuclear Structure'' Vol. I. 
 \bibitem{del04}
	 M.A. Deleplanque {\it et al.}, Phys. Rev. C {\bf 69}, 044309 (2004).	  
 \bibitem{shi05}
	 Y.R. Shimizu {\it et al.}, Phys. Rev. C {\bf 72}, 014306 (2005).	  
 \bibitem{and81}
	 C.G. Anderson {\it et al.}, Nucl. Phys. A {\bf  361}, 147  ( 1981).	 	 
 \bibitem{wu99}
	 J.-S. Wu {\it et al.}, Phys. Rev. C {\bf 60}, 044302 (1999).	  
 \bibitem{sky3d}
	 J.A. Maruhn  {\it et al.}, arXive:1310.5946.
	 
 \bibitem{Cha97a} 
	 E. Chabanat {\it et al.}, 
	 Nucl. Phys.  A{\bf 627}, 710 (1997). 	 	 
 \bibitem{Wong73}
	 C.Y. Wong, Ann. Phys. {\bf 77}, 279 (1973).
\end{thebibliography}
\end{document}